\documentclass[12pt]{article}
\usepackage{newtxtext,newtxmath}

\usepackage{graphicx}

\usepackage[letterpaper,margin=1in]{geometry}

\renewenvironment{abstract}
	{\quotation}
	{\endquotation}

\date{}

\makeatletter
\renewcommand{\fnum@figure}{\textbf{Figure \thefigure}}
\renewcommand{\fnum@table}{\textbf{Table \thetable}}
\makeatother

\usepackage{scicite}

\usepackage{xurl}

\usepackage{siunitx}
\usepackage{subfig}
\usepackage{nameref}
\makeatletter
\providecommand\@safe@activesfalse{}
\providecommand\@safe@activestrue{}
\makeatother

\def\scititle{Crowd Dynamics in Historical Perspective: Reframing the Amritsar Massacre through Agent-Based Modelling and Social Psychology
}

\graphicspath{{figs/}}

\title{\bfseries \boldmath \scititle}

\author{
	Mohcine Chraibi$^{1\ast\dagger}$,
	Krisztina Konya$^{1\dagger}$,
	Ezel Üsten$^{1}$\and
	\small$^{1}$Civil Safety Research, Forschungszentrum Jülich GmbH,  Jülich \& 52428, Germany.\and
	\small$^\ast$Corresponding author. Email: m.chraibi@fz-juelich.de\and
	\small$^\dagger$These authors contributed equally to this work.
}

\begin{document}
\maketitle

\begin{abstract} \bfseries \boldmath
Crowds have long held a paradoxical place in the human imagination, feared for their destructive potential yet essential for collective expression. This tension was tragically manifested in the 1919 Jallianwala Bagh massacre, when British colonial troops opened fire on a peaceful gathering in Amritsar, India. Although officially 379 deaths were recorded, eyewitnesses and historians have long challenged this figure.
With this study, we critically revisit the events through the lens of the specific role of the crowd as a phenomenon, both regarding the physical and the socio-psychological dynamics. We show that even under conservative physical assumptions — moderate shooting cadence, crowd-shielding, and constrained escape routes — our agent-based simulations consistently yield fatality estimates well above the official death count. On the socio-psychological front, we explore how early 20th-century discourses, influenced by Le Bon’s theory of crowd psychology, constructed the crowd as an inherently irrational and threatening entity, thus providing a rationale for the application of excessive force.
Our findings show that acknowledging the socio-cultural construction of crowds as a relevant factor in how state power engage with and respond to collective gatherings brings to light contemporary parallels and the risks posed by their rhetorical framing. Furthermore, this study highlights the importance of interdisciplinary modelling for both historical accountability and current crowd safety, particularly in an era of growing political unrest, surveillance, and militarised crowd policing.
\end{abstract}

\noindent
This study introduces an interdisciplinary framework for re-examining the 1919 Jallianwala Bagh massacre by combining agent-based simulation with critical social psychological analysis. Simulations grounded in conservative physical and historical constraints produce casualty estimates substantially higher than official records, offering a new quantitative perspective on the event. The work also examines how early theories of crowd psychology contributed to narratives that dehumanised collective gatherings and legitimised state violence. By linking historical analysis with computational modelling, the study demonstrates how simulation can serve as a tool for historical investigation and accountability while highlighting the enduring influence of crowd narratives on contemporary responses to protest and mass gatherings.








\section*{Introduction}

\textit{``There was no reason to further parley with the \textbf{mob}, evidently they were there to defy the arm of the law $\cdots$''} \cite[p.~46]{hunter-disorders_1920}

Although it was deemed ``evident'' that the crowd intended to defy British colonial law in the events leading up to the shooting, the Jallianwala Bagh massacre, commonly known as the Amritsar massacre, remains one of the most significant and deeply controversial events in colonial history. On 13 April 1919, a large, unarmed gathering assembled at Jallianwala Bagh in Amritsar, Punjab. Brigadier-General Dyer, having issued a proclamation earlier that morning prohibiting all meetings, proceeded to the enclosed garden with a contingent of troops and, without warning, ordered them to open fire on the crowd. The shooting continued for approximately ten minutes, killing hundreds and injuring many more. In addition, no medical aid was given to the wounded in the aftermath.

Historical accounts and official reports offer differing perspectives on the events and their aftermath. British authorities initially estimated the death toll at around 200, whereas estimates from the Indian side suggest a substantially higher number of casualties. Mahatma Gandhi, for instance, claimed that approximately 1,500 people lost their lives in the incident \cite[p.~219]{Wagner2019}. The Disorders Inquiry Committee Report (commonly known as the Hunter Committee), established to investigate the disturbances, provides a detailed account of the events, including the broader context of unrest leading up to the massacre, Dyer's actions, and the subsequent inquiries. The report highlights that Dyer had issued a proclamation stating that gatherings of more than four people would be dispersed by force and he was aware that a public meeting was planned at Jallianwala Bagh. Dyer himself claimed that the crowd had assembled in defiance of his orders.

The events leading up to the Jallianwala Bagh meeting were marked by political unrest and widespread protests against the Rowlatt Act, which granted authorities extensive powers to arrest and detain individuals without trial, with the aim of suppressing nationalist and revolutionary movements in India. Amritsar had been a major centre of this agitation, and the deportation of key movement leaders on April 10th further inflamed tensions in the city, leading to widespread unrest and ultimately triggering the events of April 13. Despite prohibitory orders, the meeting at Jallianwala Bagh proceeded, attracting a large number of people, some possibly unaware of the ban and others perhaps intending to witness a \textit{tamasha} or spectacle.

This article revisits the events of the massacre from the perspective of the specific role played by the crowd. First, we reconstruct the societal discourse on crowds around the turn of the century and examine how these ideas are reflected in the personal account of the commanding British general regarding the events resulting in April 13, 1919. The second approach employs an agent-based simulation to reconstruct the massacre using a mathematical model and to identify crowd-specific aspects that are beyond a mere individualistic modelling approach. By combining interdisciplinary perspectives, this study aims to reassess the tragedy with scientific rigour, focussing on the dominant attitudes toward crowds at the time, as well as the role of specific characteristics of the massacre such as reported firing rates, the dynamics of collapse among the crowd, and the impact of exit congestion and flow control. This integrated approach can provide new insights into the dynamics and death toll of the event, potentially challenging existing historical narratives. 

Furthermore, beyond reassessing this specific historical event, the paper aims to introduce a broader discourse on the applicability of crowd simulations to understand crowd dynamics in cases where crowds are the direct targets. Contemporary crowd science has largely focused on high-density challenges in the context of orderly movement, such as evacuation or ingress/egress optimisation. However, as  recent attacks on public gatherings \cite{euronews2024christmas} have shown, crowds can also become explicit targets of violence, leading to distinct behavioural and organisational dynamics. In such scenarios, individuals self-organise in response to a \textbf{centralized threat} (e.g., a bomb or active shooter) and must adapt to an ever-changing environment (where casualties may become physical obstructions or even unintended shields), producing patterns that current simulation frameworks may be ill-equipped to capture. Building on the perspectives introduced in this study, we aim to demonstrate how these methods can be used to both retrospectively reconstruct and verify events of this nature , and to prospectively identify spatial and organizational vulnerabilities in future public gatherings. In doing so, simulations may not only serve explanatory purposes, but also challenge narratives that downplay the severity of violent crowd incidents. The Jallianwala Bagh massacre is one such case, where this approach offers a means to interrogate historical records and provide a scientifically grounded estimate of crowd behaviour and casualties, potentially acting as a forensic counterweight to narrative minimisation.

\begin{figure}[htbp]
    \centering
    \includegraphics[width=1\linewidth]{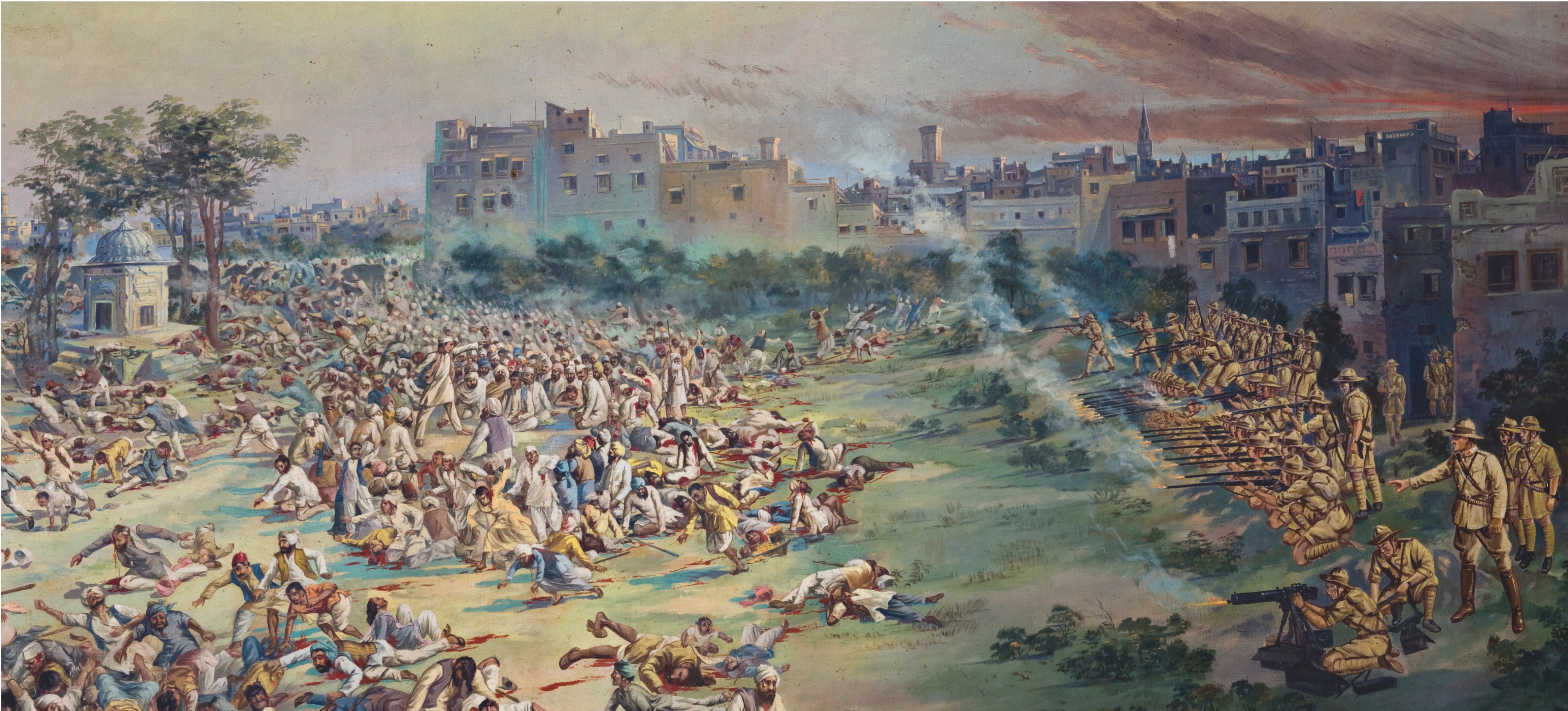}
    \caption{ An artistic representation of the Jallianwala Bagh massacre, showing British troops firing into a densely packed and unarmed crowd. This image, illustrates the one-sided nature of the violence, the absence of cover or escape, and the magnitude of civilian suffering. Reproduced with permission of Kim A. Wagner.}
    \label{fig:massacre-painting}
\end{figure}

To situate our simulation within the lived and remembered experience of the Jallianwala Bagh massacre, we include a historical painting that depicts the event from the perspective of Indian memory. 
Figure~\ref{fig:massacre-painting} offers a poignant visual account of the violence unleashed by British troops on the unarmed crowd. Although artistic, this rendering reflects key elements corroborated by survivor testimonies and military records: unidirectional volley fire, dense crowd composition, and the absence of safe exits. The image conveys the terror and chaos that simulations seek to represent in an abstract form. 

\section*{Historical background}
\label{sec:historical-background}

The tragic events examined in this article took place on 13 April 1919 at Jallianwala Bagh, a public park in Amritsar commonly used for recreation. On the day of the massacre, a protest group had called for a political gathering at the site; however, most visitors to the Bagh were unaware of the purpose of the meeting. According to Kim A. Wagner's previous historical research, Brigadier-General Reginald Dyer, already in a highly agitated state, interpreted the gathering as a defiance of recently imposed prohibitions on public assembly and a threat to British authority. Just three days earlier, tensions had escalated in the city due to clashes between demonstrators and British authorities (for detailed information, see \cite{Wagner2019}). The massacre took place around noon, after Dyer and his troops arrived in the city at Bagh. He entered the Bagh with fifty soldiers and, without issuing any warning, ordered ten minutes of sustained gunfire on the crowd. 

Wagner identified several factors that contributed to the massacre, including a profound overestimation of the threat posed by the local population. Dyer's troops, already on edge from previous encounters during their march through the city, had made arbitrary arrests before arriving at the park \cite[p.~113]{Wagner2019}. Despite the largely non-political character of the crowd, Dyer interpreted the gathering as a direct provocation. Inaccurate reports further fed this perception, portraying the gathering as a part of a wider uprising, extending beyond the crowd assembled in the Bagh. Dyer perceived himself as upholding imperial authority and, as Wagner notes, ``\textit{the perceived need to maintain British prestige and saving face at all costs thus imbued Dyer's actions with a crucially performative function}'' \cite[p.~148]{Wagner2019}. The absence of local police, who refrained from intervening directly, further amplified the sense of urgency and vulnerability of the military \cite[p.~120]{Wagner2019}.

Dyer had strategically split his troops to secure the city gates and instructed an officer to act if his unit did not return by 2 p.m., reflecting his sense of personal jeopardy. He later claimed that he ``knew'' that the crowd was composed of the same ``mobs'' responsible for previous violence, effectively projecting past incidents onto the present situation: ``\textit{Dyer was, therefore, not reacting to the actual crowd in front of him as much as to what he imagined that crowd to be}'' \cite[p.~148]{Wagner2019}. 

An important influencing factor was the perceived appearance of the crowd. Eyewitnesses uniformly describe the gathering as overwhelmingly large \cite[p.~117]{Wagner2019}. According to Wagner's account of Dyer: ``\textit{Dyer, who had never been to Jallianwala Bagh before, was overwhelmed by the sheer size of the gathering that he had walked in on}'' \cite[p.~121]{Wagner2019}. This perception culminated in a dehumanising view of the crowd as a unified, threatening entity. Wagner concludes, `\textit{the firing $\cdots$ had instead become a pure spectacle of brute force in which the `rebels' were perceived as an undifferentiated mass}'' \cite[p.~151]{Wagner2019}.

In considering the causes of the massacre, it is therefore essential not only to consider strategic, political, and psychological dimensions, but also to recognise the cultural and epistemic frameworks that shaped perceptions of the crowd. Frameworks that, in turn, helped to justify such violence. In particular, we emphasise the importance of understanding how dominant views of crowds as inherently dangerous, irrational, and dehumanised collectives, a perspective that may have contributed significantly to the unfolding of this atrocity, have shaped historical (and contemporary) reactions.

\subsection*{Crowds in historical context}
The phenomenon of the crowd was one of the most inspiring social phenomena of the late 19th and early 20th centuries. Rapid industrialisation has drawn large segments of the rural population to urban centres, resulting in unprecedented concentrations of people. At the same time, crowds became increasingly a symbol and instrument of political power in the struggle for social transformation. Iconic images of the French Revolution with beheading, and especially images of the Paris Commune with violent mass actions, took on a symbolic character, embodying the claims to power of emerging new social forces that challenged traditional power relations. This perception is best illustrated by the frequently cited eyewitness report of Georges Clemenceau about the Paris Commune atrocities: ``\textit{All were shrieking like \textbf{wild beasts} without realising what they were doing. I observed then that pathological phenomenon which might be called \textbf{blood lust}. A breath of \textbf{madness} seemed to have passed over this mob $\cdot$}'' \cite[p.~2]{nye_origins_1975}. The heightened emotional arousal causes all nuance to fade, allowing perception to be overtaken by the image of the mob, a term marked by negative connotation and associated with the irrationality of the ``madding crowd.'' Clemenceau's wording reflects this mindset: rather than portraying the crowd as a gathering of individuals, it is framed as a volatile and threatening mob, effectively suppressing differentiated observation and reinforcing a dehumanising narrative.

The negative connotation associated with mass behaviour was also reflected in academic discourse on crowds at the turn of the century. From Hyppolyte Taine, acknowledged as the first crowd psychologist \cite{nye_origins_1975}, to Gabriel Tarde, Scipio Sighele, and Gustave Le Bon \cite{LeBon1896, palano_science_2025, borch_crowd_2021, barrows_distorting_1981}, early studies of crowd behaviour were framed within paradigms ranging from criminalisation (both Tarde and Sighele were criminologists) to psychological incapacitation. Le Bon's approach was to refute a generalised criminal orientation of crowds and to provide an alternative explanation for the destructive power of crowds. In doing so, he drew on the assumption that crowds form a psychological entity based on the same emotional attitude and on the unconscious character of behaviour in crowds. This standardisation is achieved through mechanisms of contagion that bypass cognitive elements (i.e., being rational) and, without exception (because they are associated with archaic characteristics), capture all those present in a crowd. The image of the criminal masses, which had characterised the image of gatherings of crowds in public spaces since the French Revolution, was replaced by that of unconsciously destructive masses. The emotional and unconscious crowd entity was contrasted with the rationally acting individual. Yet, this opposition continued to reflect a dehumanising perspective, as it implied that individuals were psychologically altered and regressed into a primitive or uncivilized condition when they join the entity of a crowd \cite{palano_science_2025}. The masses were thus assumed to be an irrational power that can and will be used as a counterweight against the bourgeois values of education and intellect. 
As historical sources show, Le Bon's views were known far beyond French borders. The work saw notable circulation in Great Britain, with 12 editions published between 1896 and 1920 \cite[p.~1]{jupp_introduction_2000} and went on to shape academic discourse on crowd behaviour over the following decades \cite{moscovici_age_1985, nye_origins_1975, barrows_distorting_1981}, an influence that remains discernible even today \cite{thonhauser_critique_2022}.

\subsection*{A crowd psychology perspective on the massacre}
The views discussed in the previous section can be observed not only in the progression of events in Amritsar but also in the way they were reported. This is particularly clear in General Dyer's interviews and in the language of the Hunter report, which frequently portrayed the crowd as inherently aggressive, volatile, and irrational. The repeated use of the term ``mob'' throughout the report reflects a broader discursive strategy, one that ascribed hostility and disorder to collective civilian action, thereby justifying violent repression.
\textit{``Before dealing with other outrages committed by the \textbf{mob}, it is necessary to make special comment $\cdots$''} \cite[p.~36]{hunter-disorders_1920}.
The Hunter Report further reinforced this narrative by drawing a sharp distinction between the so-called ``loyal'' citizens and those who resisted colonial authority. Loyalty was equated with rationality, while defiance was framed as irrational and dangerous.
\textit{``Other efforts by \textbf{sane} and loyal citizens inside the city on that day we have no doubt there were.''} \cite[p.~40]{hunter-disorders_1920}

Regarding Dyer's actions and the rationale behind them, his primary motivation appears to have been asserting dominance: \textit{``\textbf{The crowd}, in complete defiance of my orders, forced my hand, and it was my duty to vindicate authority.''} \cite[p.~116]{Wagner2019}. He repeatedly emphasised the need to respond to any perceived challenge to his authority, fearing that failure to act decisively would undermine both his own reputation and the prestige of British rule: \textit{``That a counter-proclamation had been issued behind me, that the rumour had been set going that my action was mere pretence, and that I dared not fire.''} \cite[p.~114]{Wagner2019}. He further described the gathering as equivalent to an act of rebellion: \textit{``The whole Punjab had its eyes on Amritsar, and the assembly of the crowd that afternoon was for all practical purposes a \textbf{declaration of war} by leaders whose hope and belief was that I should fail to take up the challenge.''} \cite[p.~122]{Wagner2019}. In stark contrast, local organisers had chosen the Bagh precisely because it was viewed as a safe, unnoticeable location. As Wagner notes, \textit{``A meeting in that place could not reasonably be interpreted as a provocative or aggressive action – or so it was assumed.''} \cite[p.~117]{Wagner2019}.

The power dynamics underlying Dyer's motivations reflect the influence of Le Bon’s perspective on crowds, not explicitly cited, but undoubtedly recognized. Although Dyer had issued several proclamations the day before the massacre, he responded with drastic force only to the large assembly at Jallianwala Bagh. His enforcement of other measures, such as curfews and movement restrictions, was largely limited to a symbolic military procession through parts of the city. These efforts were inconsistently applied, poorly communicated, and ultimately ineffective. The idea of ``sending a strong message'' was based on confronting a powerful entity — in this case, a large, unsanctioned gathering — which, by its very nature, was perceived as demanding immediate and decisive suppression.

The logic follows a similar trajectory to that of Le Bon: for him, the crowd represented a mindless and dangerous mass, threatening the social and political order of the elite class and, by extension, his own existence. Dyer, similarly, perceived the crowd not as a gathering of civilians but as an aggressive and potentially murderous mob (a framing reinforced by the colonial discourse of the time, which cast nationalist and revolutionary movements not as political expressions but as existential threats to imperial authority). However, this narrative directly contradicted the available evidence: the organisers had chosen Jallianwala Bagh precisely because it was not a provocative space; the crowd included many children; and the city-wide announcement of martial law proclamations was inconsistent at best. Even Dyer later acknowledged this uncertainty, admitting: \textit{``I confess I do not know how far we had penetrated into the city. I do not know the city very well $\cdots$ There may have been a good many who had not heard the Proclamation.''} \cite[p.~112]{Wagner2019}. 

Furthermore, the gathering took place during a major religious festival, attracting many individuals from outside the city who may not have been fully aware of the political situation taking place in Amritsar. The crowd was therefore not homogeneous in its purpose. Dyer perceived the assembly not as a diverse group of individuals but as a singular, defiant entity challenging British authority. This perspective did not account for the varied reasons individuals had for attending the gathering, including participation in the Baisakhi festival and peaceful protest. 

\section*{Model definition}
\label{sec:results}


In the previous sections we highlight the particular mindset surrounding crowds, which can be regarded as an additional socio-cultural factor influencing how conflict-laden political situations are approached and managed.
In the model, we focus on the dynamics triggered by the gunfire. Our aim is to reconstruct possible scenarios of the massacre. In doing so, we demonstrate how spatial conditions contributed to the crowding effects during gunfire and how these crowding effects may have influenced the resulting deaths and injuries.

To investigate the spatial and collective effects of crowd exposure under fire, we developed a physically grounded agent-based simulation of the Jallianwala Bagh massacre scenario. The aim was to test whether plausible assumptions about crowd density, exit constraints, and targeting strategies could account for a scale of fatalities beyond official reports. The simulation structure, parameter assumptions, and key outcomes are presented below.


We simulate pedestrian behaviour using a first-order collision-free speed model~\cite{Tordeux2016}, in which each agent's motion follows:
\begin{equation}
    \dot{x}_i = V(x_i, x_j, \dots) \cdot \vec{e}_i(x_i, x_j, \dots),
\end{equation}
where $x_i$ is the position of pedestrian $i$, $V$ is a speed function based on spacing to the neighbors, and $\vec{e}_i$ is the normalized direction of motion. 
Pedestrians begin with an initial preferred speed $v_0 = \SI{3}{\meter\per\second}$, representing running escape behaviour.

Refer to \nameref{sec:simulation} for the detailed breakdown of the simulation results.

\subsection*{Collapse dynamics and density effects}
To model the likelihood of pedestrian collapse under gunfire, we define a survival probability $p(x,t)$ that depends on both spatial location and exposure time. 
This probability reflects the cumulative threat experienced by a pedestrian at position $x\in \mathbb{R^2}$ and time $t$:
\begin{equation}
    p(x,t) = r_\text{space}(x) \cdot r_\text{time}(t).
    \label{eq:survival_prob}
\end{equation}
Here,  \(r_\text{space}\) quantifies the spatial risk based on proximity to gunfire, while  \(r_\text{time}\) accounts for the temporal effect of prolonged exposure.

We assume that bullets are fired continuously and independently by a vertical line of shooters stationed along the left boundary of the area. Each pedestrian is thus exposed to a cumulative spatial risk resulting from all shooters in this line.
The total spatial exposure is computed as:
\begin{equation}
    r_\text{space}(x) = 1 - \frac{R(x)}{R_\text{max}},\; R(x) = \sum_{i=1}^{N} \frac{1}{1+\frac{\|x-s_i\|^2}{\sigma^2}},
    \label{eq:rawrisk}
\end{equation}
where \(s_i\) denotes the position of the \(i\)-th shooter along the firing line, \(N\) is the total number of shooters, and \(\sigma\) controls the spatial decay of bullet lethality. According to historical evidence \cite{Wagner2019} $N=50$.
In order to avoid regions that are completely safe or inevitably lethal, the spatial risk factor is restricted to the interval [0.05, 0.95]. 

\begin{figure*}[htbp]
\centering

\subfloat[$t = \SI{0}{\second}$]{%
\includegraphics[width=0.4\textwidth]{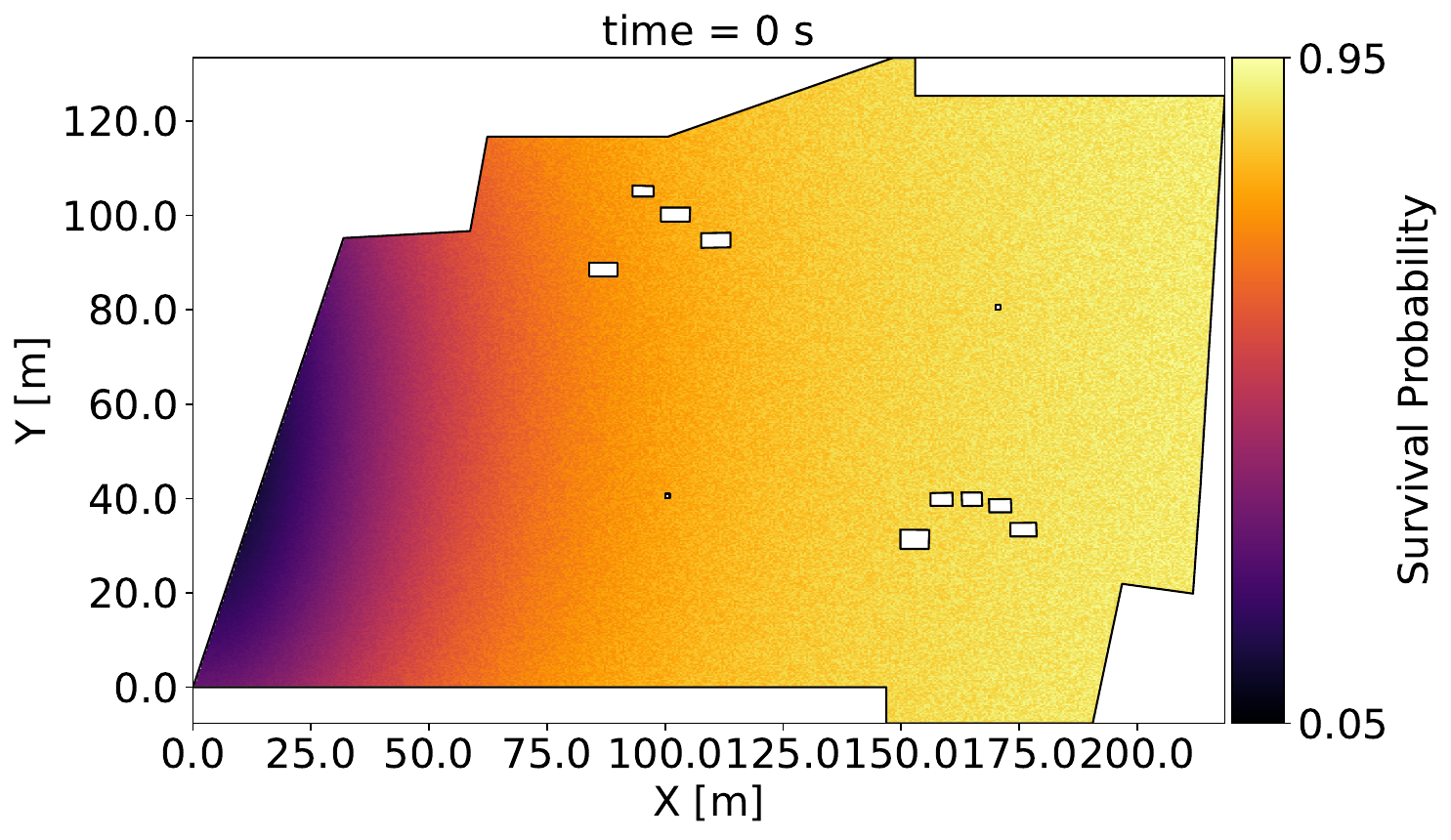}}
\hfill
\subfloat[$t = \SI{200}{\second}$]{%
\includegraphics[width=0.4\textwidth]{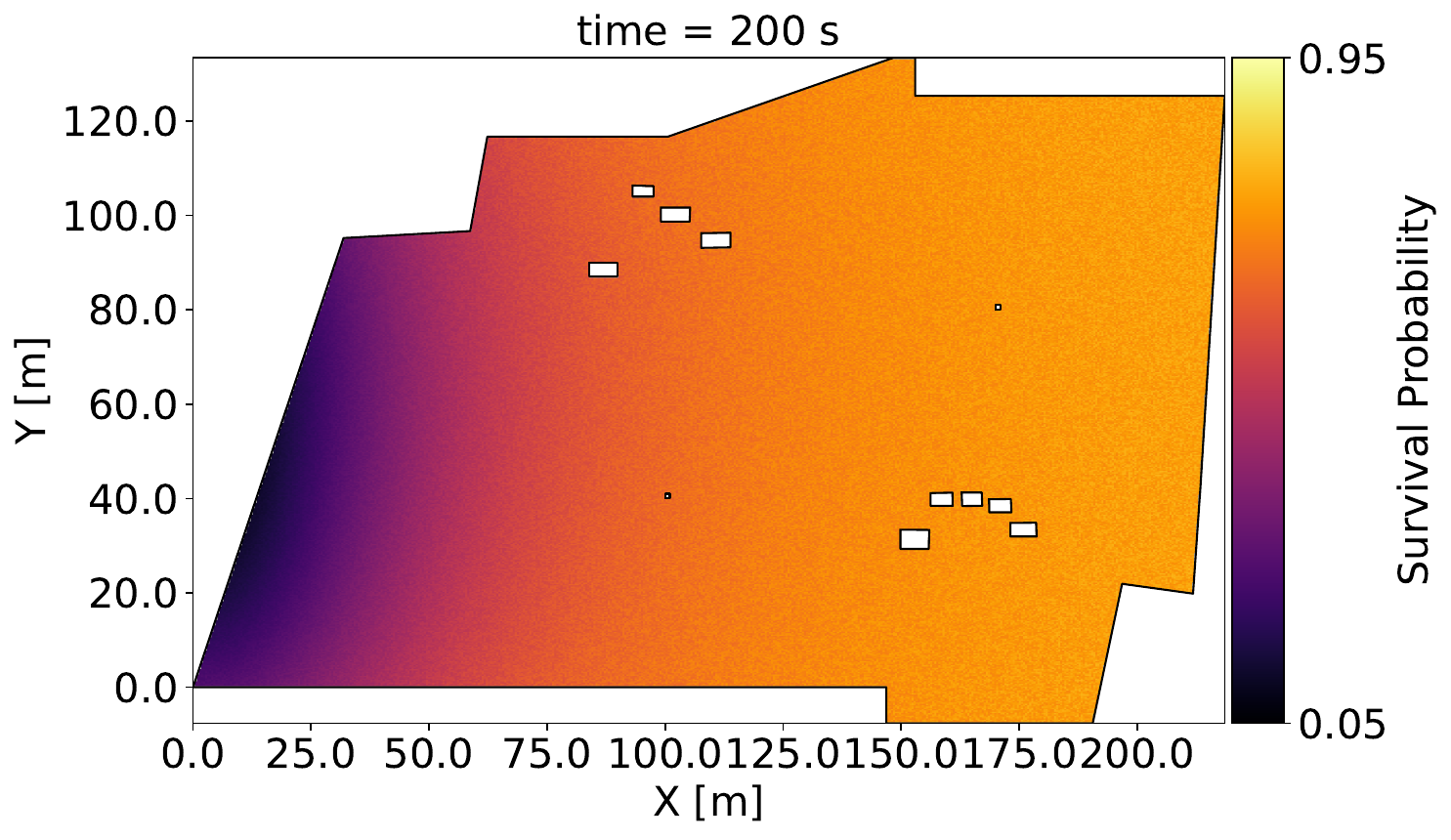}}

\vspace{0.5em}

\subfloat[$t = \SI{400}{\second}$]{%
\includegraphics[width=0.4\textwidth]{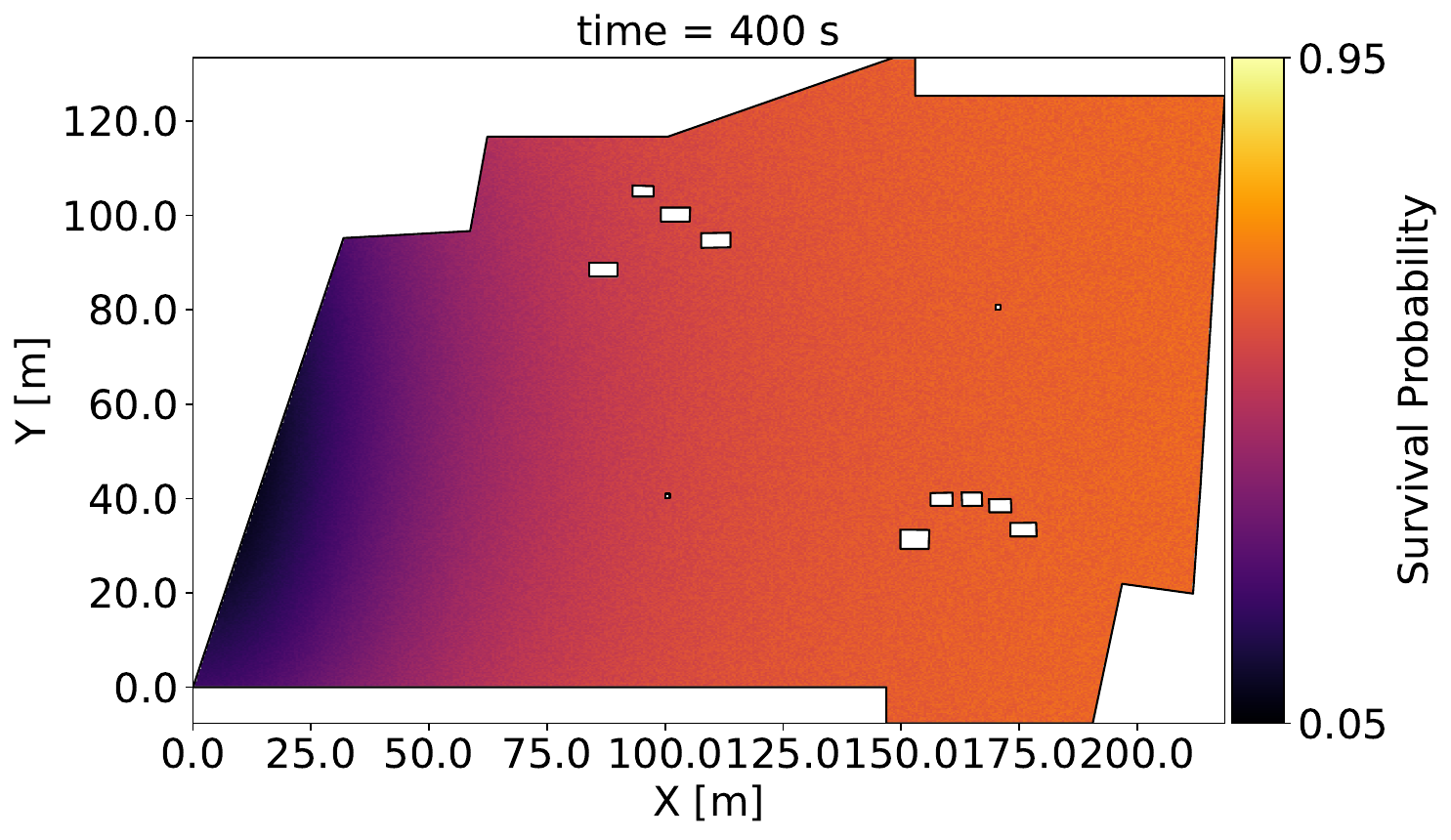}}
\hfill
\subfloat[$t = \SI{600}{\second}$]{%
\includegraphics[width=0.4\textwidth]{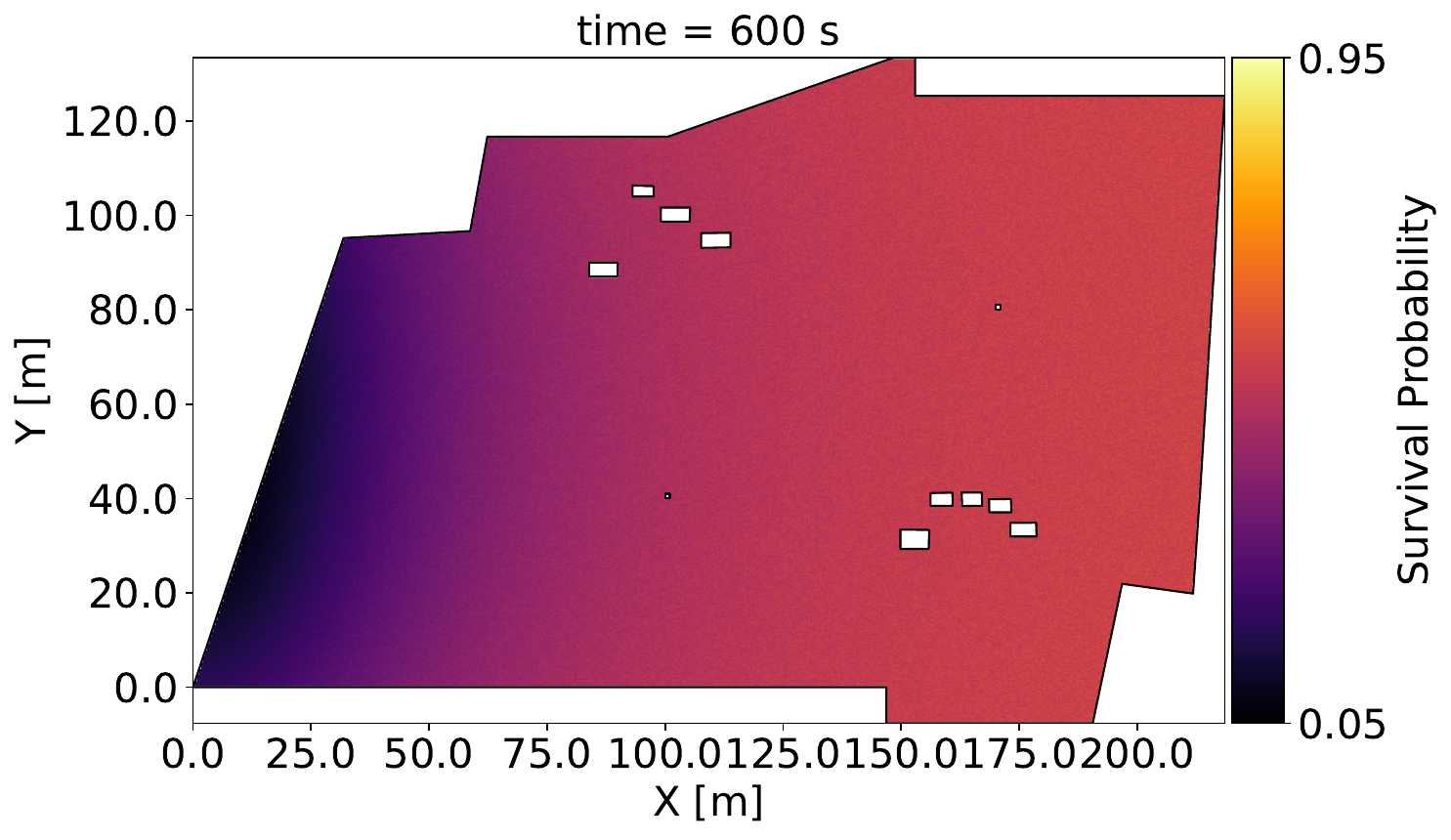}}

\caption{Spatiotemporal survival probability fields at different time points according to Equation~\ref{eq:survival_prob}. Survival probability decays over time due to prolonged exposure, while spatial variation reflects the risk gradient from the firing line located at the left edge of the area.}
\label{fig:spatiotemporal_survival}

\end{figure*}




\textbf{Temporal exposure model.} We model temporal exposure using an exponential decay function $r_\text{time}(t) = \exp(-\lambda_i t/T)$, where $T = \SI{600}{\second}$ corresponds to the duration of the shooting event reported in \cite{Wagner2019}. To introduce heterogeneity and prevent rigid thresholding, each agent $i$ has an individual decay rate $\lambda_i \sim \mathcal{U}(\lambda - 0.1, \lambda + 0.1)$, adding population-level variability in survival probability while maintaining model robustness.

\begin{equation}
    r_\text{time}(t) = \exp\left(-\lambda \frac{t}{T}\right),
    \label{eq:rtime}
    \end{equation}

The parameter \( \lambda \) governs how quickly the survival probability decreases. 
In our simulations, we select conservative values for \( \lambda \) to avoid overly pessimistic scenarios in which agents die ``unrealistically'' quickly. 
Figure~\ref{fig:spatiotemporal_survival} shows for $\lambda=0.5$ the change in survival probability for different times throughout the whole scenario.

\subsection*{Crowd shielding effect}
\label{sec:crowd_shielding}
In densely packed environments such as Jallianwala Bagh, exposure to danger varied significantly among individuals. 
We can assume that physical positioning created natural protection. 
Some people were protected by others, blocked from direct fire lines, or sheltered by structural obstacles and the density of the crowd.
However, historical evidence \cite{Wagner2019} indicates that soldiers may have deliberately targeted the densest concentrations of people. This targeting strategy would reverse the protective effect of crowds, transforming high-density areas from refuges into zones of increased danger.

To capture this \textbf{dual nature of crowd effects}, we implement a crowd shielding mechanism that models two competing dynamics: the protective benefit of being surrounded by others versus the increased target risk of dense clusters. 

 The collapse probability becomes:
\begin{equation}
    P_\text{collapse}(x, t) = 1 -  \Big(p(x,t)\cdot  (1+\gamma[\alpha s+(1-\alpha)(1-s)]) \Big),
    \label{eq:collapse}
\end{equation}
where \(s =  \min\left(1, \frac{n}{n_{\text{max}}}\right) \in [0,1]\) represents the local shielding level,  and \(\gamma\) controls the strength of this effect.
Here \(n\) is the number of neighbouring agents within \SI{1.5}{\meter}, and \(n_{\text{max}} = 12\) defines the saturation threshold.

\noindent The parameter $\alpha\in [0,1]$  determines which mechanism dominates. 
At $\alpha=1$, crowds provide physical protection, agents surrounded by others face reduced risk. 
Whereas at $\alpha=0$, crowds become a liability, dense clusters attract concentrated fire, increasing the  probability of collapse. The intermediate values of $\alpha$ blend these opposing effects.

Higher values of the \(s\) values indicate a stronger local density, which either enhances protection or increases targeting risk depending on the value of $\alpha$.
This flexible formulation captures the ambiguity of crowd effects in violent scenarios, where physical shielding and exposure amplification can coexist.

Although we currently introduce a linear scaling parameter $\gamma$ to control the strength of the shielding effect, future work could explore embedding this directly into a nonlinear function of local density. This would eliminate the need for an explicit scaling parameter and offer a more emergent representation of the shielding dynamics.

The formulation in Equation~\ref{eq:collapse} captures how the risk of collapse arises from the interaction of spatial exposure, temporal duration, and local crowd configuration. Agents in high-risk situations, that is, those positioned close to the firing line, exposed for extended periods, and  isolated (when $\alpha$ is high) or embedded in densely packed clusters (when $\alpha$ is low), face elevated collapse probabilities.
Conversely, agents benefit from protective spatial positioning, brief exposure, and favourable crowd dynamics, either through shielding provided by nearby individuals or by avoiding areas of high concentration, depending on the tactical scenario governed by $\alpha$.

In regular update steps (see Section \nameref{sec:gunfire}), agents collapse stochastically with probability \(P_\text{collapse}(x,t)\). 
Collapsed agents remain immobile and act as static obstacles, further constraining local movement. This process models both fatalities and incapacitations due to injury or immobilisation.

\subsection*{Geometry and exit flow constraints}

Jallianwala Bagh was a walled courtyard enclosed by the rear sides of residential buildings and high brick walls. The enclosed area was irregularly shaped, approximately $\SI{182}{\meter}$ long and $\SI{137}{\meter}$ wide, and resembled a distorted trapezoid.
According to Indian journalist Madan Mohan Malaviya, who visited the site shortly after the massacre, the exits were severely restricted. 
\begin{quote}
    ``There are no proper doors, but there are five narrow openings or shabby alleyways that serve as entrances and exits.'' \cite{Malaviya1919}
\end{quote}

To reflect these constraints in our simulation, we introduce a flow limiter at each exit, restricting it to $20\%$ of its theoretical maximum capacity. This adjustment accounts for the severe bottlenecks and physical obstructions that would have impaired evacuation.

\begin{figure}[htbp]
    \centering
    \includegraphics[width=0.9\linewidth]{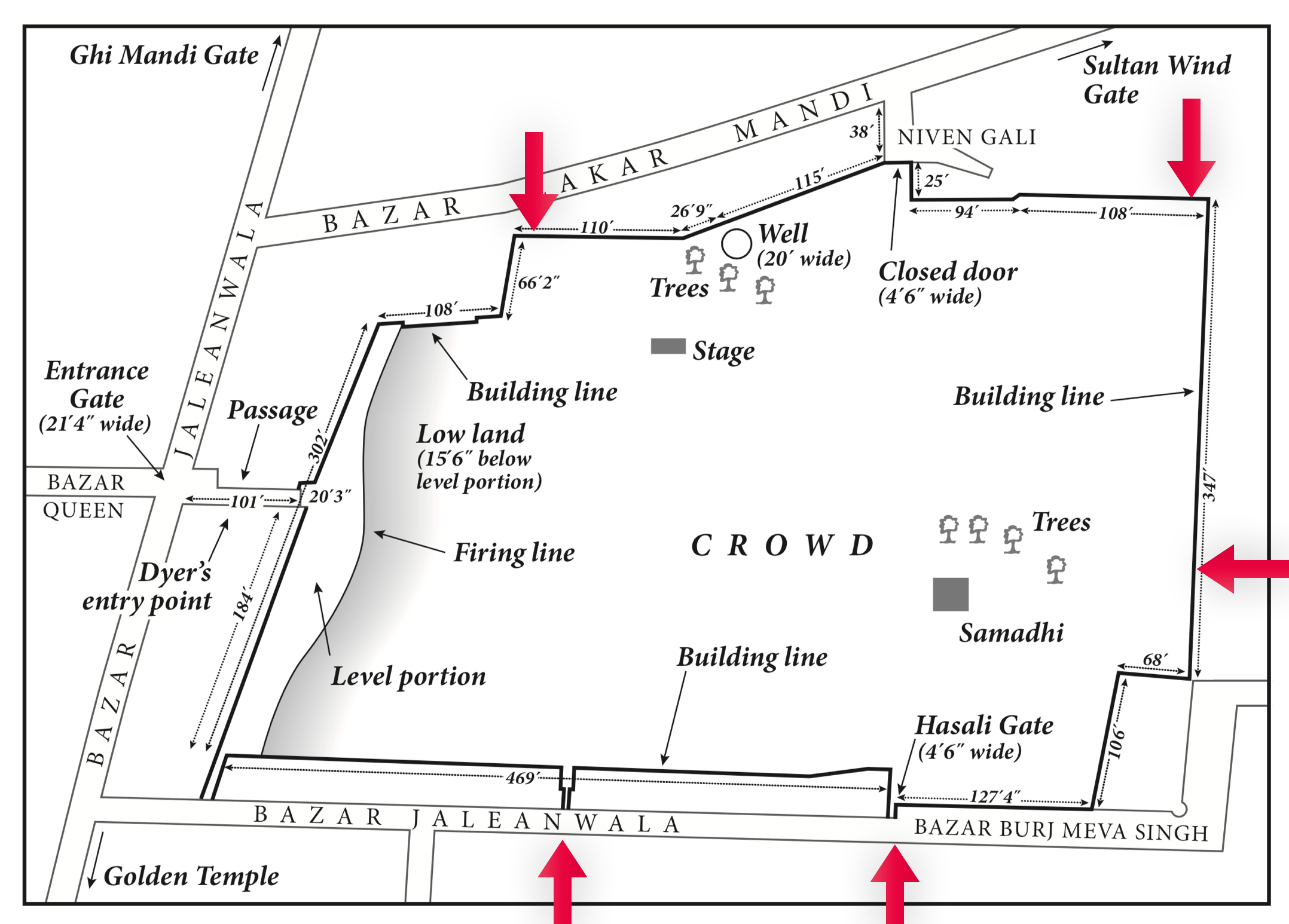}
    \caption{Reconstruction of the Jallianwala Bagh geometry showing the enclosing walls, exit points (red arrows), internal features (trees, well, and stage), and Dyer's entry and firing line. Distances are marked in feet and inches, following the historical map. Reproduced with permission of Kim A. Wagner.}
    \label{fig:geometry}
\end{figure}

Figure~\ref{fig:geometry} shows a reconstructed layout of the Bagh. 
The five exit points are marked with red arrows. We chose these locations because of the relatively lower level of the walls there, so we can assume that they could be used to exit Bagh.
The stage (used during the protest gathering) is located near the centre of the open area. 
A well and trees offered limited vertical obstruction, while the presence of low land near the entrance created further spatial variation. In the simulation, we approximate all these objects with simple polygons acting as obstacles.
The diagram also marks the firing line where the soldiers took their position and blocked the main passage (left).

To model realistic decision-making under uncertainty, we can assume that pedestrians in the Bagh are not always running towards the nearest exit deterministically. Instead, we assign a selection probability to each exit based on its distance using Equation~\ref{eq:exit_prob}
\begin{equation}
P_{\text{ exit}_i} = \frac{(d_i + \varepsilon)^{-\beta}}{\sum_j (d_j + \varepsilon)^{-\beta}},
\label{eq:exit_prob}
\end{equation}
where \( d_i \) is the distance to exit \( i \), \( \beta \) is a proximity bias parameter, and \( \varepsilon = 10^{-6} \) is a small constant added for numerical stability.
The term \( \varepsilon \) ensures well-defined behaviour even when a pedestrian is located exactly at an exit (\( d_i = 0 \)), avoiding division by zero and enabling smooth probability computation.



\subsection*{Temporal resolution of gunfire effects}
\label{sec:gunfire}
Official records indicate that approximately $1{,}650$ bullets were fired in ten minutes by 50 soldiers at Jallianwala Bagh~\cite{Wagner2019}. This translates to an average of just over 3 rounds per soldier per minute, far below the technically possible firing rate of 15–30 rounds per minute for trained riflemen. Contemporary eyewitness accounts and military records suggest that the shooting was not executed in rapid volleys but rather as deliberate sustained fire aimed broadly at the crowd.

In light of this, our simulation distributes the firing exposure evenly over the course of the event, assuming a slow and continuous fire pattern. 
Although the firing was deliberate and not organised in volleys, we choose a \SI{10}{\second} update interval to recalculate the collapse probabilities. This timescale provides a computationally tractable and behaviourally plausible resolution to simulate the evolution of local risk, crowd movement, and exposure accumulation. It allows agents' survival chances to respond dynamically to their changing surroundings while avoiding unrealistic instantaneous mass collapse.

In our model, agent collapse represents incapacitation, either due to fatal injury or non-lethal trauma. Since we do not distinguish between death and injury, we refrain from enforcing a strict upper bound on the total number of collapses. 
The known figure of 1\,650 rounds serves as a physical constraint reference, but not as a hard limit on outcomes, as a single bullet may incapacitate more than one person or be reflected in crowd dynamics beyond simple lethality.

\subsection*{Interpretive summary: dual role of crowding}
One of the most revealing insights from our model is the dual role of crowd density as both protective and hazardous. This dynamic, governed by the shielding parameter $\alpha$, underlines a central tension in collective behaviour under targeted violence:

A particularly important aspect emerging from these results is the dual role of the crowd itself as both a protective barrier and a source of increased exposure. 
In our model, this is captured by the shielding parameter $\alpha \in [0,1]$ (Equation~\ref{eq:collapse}, which determines whether the crowding effect predominantly shields individuals ($\alpha=1$) or amplifies their vulnerability ($\alpha=0$). 
At high $\alpha$, individuals surrounded by others are partially protected from direct gunfire, illustrating how physical clustering can reduce immediate risk. Conversely, at low $\alpha$, the crowd acts as a liability: dense clusters attract concentrated fire, increasing the probability that bystanders behind initial victims are also struck. 
This means that being embedded in a dense group can raise the probability of becoming a secondary or tertiary victim, leading to a cascading failure mechanism. 

This insight underscores a key distinction between modelling mass-casualty crowd events and modelling individual risk at an aggregate level. 
It is not sufficient to treat individuals as isolated units; the emergent dynamics of crowding, movement constraints, and spatial clustering fundamentally alter exposure patterns. 
By explicitly representing these interactions, our model demonstrates how collective crowd behaviour can shift the balance between protection and exposure, highlighting the need for adequate modelling of both physical and social-psychological aspects of crowd disasters due to violence.

\section*{Discussion}
\subsubsection*{Interpretation and broader implications}

Historical records of the Jallianwala Bagh massacre remain contested, with significant discrepancies between official colonial accounts and those reported by Indian witnesses and investigators. By grounding our simulation in plausible physiological and physical constraints, such as spatial exposure, firing rates, and the geometry of the Bagh, our modelling approach provides a quantitative lens to re-examine these narratives. Although we do not claim absolute accuracy in estimating the death toll, our results consistently produce collapse counts that exceed the officially recorded number of 379 deaths. This finding reinforces long-standing critiques that the colonial record may have underreported the event's scale and highlights the potential of computational methods to challenge dominant historical accounts with data-driven analysis.

Simulating historical tragedies raises important ethical questions. Modelling human suffering, even for scientific or commemorative purposes, must be done with sensitivity and respect for victims and their descendants. In this work, we are not seeking to reduce a complex human tragedy to numbers or pixels. Rather, the simulation serves as a tool to understand the mechanisms that might have contributed to high mortality, to explore how architectural confinement and crowd dynamics can escalate risk, and to foster critical reflection on how such events are remembered or obscured in historical narratives. The ethical imperative lies in transparency, in acknowledging uncertainty, and in using the model to amplify awareness rather than to sensationalise.

\subsubsection*{Limitations of the model}

Although our simulation provides insight into the spatial and temporal dynamics of deaths during the Jallianwala Bagh massacre, several limitations should be acknowledged.

First, the model does not distinguish between injury and death. All agents who collapse are treated as fatalities, without modelling post-collapse outcomes such as survival, recovery, or medical intervention. This simplification was adopted to focus on the mechanics of incapacitation under gunfire rather than the full spectrum of casualty outcomes.

Second, the model abstracts out complex human behaviours. We do not simulate panic, protective instincts (e.g., shielding children), coordinated group movements.
Instead, the probability of collapse is determined solely by each agent's spatial position, time under fire, and local density. Consequently, emergent collective dynamics, such as bottleneck turbulence or crowd collapse propagation, are not explicitly represented.

Third, we omit one of the most historically significant architectural features of the Bagh: the well. According to historical accounts \cite{Wagner2019}, approximately 120 people died after jumping into the well in a desperate attempt to escape the gunfire. Our current simulation geometry does not include this structure and therefore cannot capture clustering, refuge-seeking behaviour, or crowding effects near confined shelter. Future work incorporating the well would offer a more complete spatial reconstruction of the tragedy.

Fourth, we do not model progressive fatigue. The initial walking speed of each agent \( v_0 \) is drawn from a normal distribution and remains constant throughout the simulation. Although the probability of collapse depends on spatial and temporal exposure, we do not simulate a gradual deterioration of mobility over time. This decision prioritises interpretability and computational efficiency, but omits long-term exhaustion effects under sustained threat.

Despite these limitations, the model serves as a physically grounded and interpretable approximation of the exposure-based collapse dynamics. Future extensions could incorporate more behavioral nuance, structural detail, and physiological processes to further improve realism and explanatory power.

\subsection*{Connecting historical perspectives and simulation results}

This study aims to demonstrate that a comprehensive understanding of social and physical dynamics in crowd events (particularly in cases of mass disasters) requires consideration of the crowd as a distinct socio-cultural phenomenon. From a social psychological perspective, this involves reconstructing the socio-cultural meanings attributed to the crowd within the specific context under examination. In parallel, the simulation highlights the significance of crowding effects in shaping the outcome of the tragic incident. 

Our intention is not to challenge existing interpretations or narrow the analytical perspectives, but rather to expand the conceptual framework for understanding the underlying causes of crowd behaviour, especially in the context of violent mass events. Importantly, this approach does not negate individual accountability: Dyer's command remains his sole responsibility. Collective mindsets merely provide a framework for interpreting his actions.

As Palano \cite{palano_science_2025} emphasises, in Le Bon's theory, dehumanisation did not apply solely to the notion of the crowd as a whole. Rather, it involved attributing an unconscious and uncivilised psyche to each individual within the crowd, thereby holding every person collectively responsible for the actions of the group. In this sense, it is not only the composition of a crowd (for instance, when construed as a political adversary) that may be perceived as threatening, but also the mere physical presence of a gathering can elicit perceptions fostering subtle or implicit forms of dehumanisation — a phenomenon described by Haslam as ``infrahumanisation'' (see \cite{haslam_dehumanization_2015}). This psychological framing can function as a moral justification for acts of violence against evidently peaceful assemblies, as exemplified by the massacre at Jallianwala Bagh in 1919.

On the other hand, many simulation studies treat crowds as aggregations of individual agents, modeled independently of any overarching collective context. However, the findings of this study demonstrate that the fragmented and emerging dynamics within the attacked crowd can only be realistically represented when the crowding effects are adequately integrated. These effects are essential for modelling the event in a manner that reflects its complex reality.

Ultimately, a crowd cannot be understood merely as a homogenous mass, nor as a collection of isolated individuals. Instead, it must be approached as a relational and dynamic formation: shaped by the interactions between individuals, the meanings attributed to the crowd both internally and externally, the social context of the gathering and the structural constraints that emerge from these interconnections. A comprehensive understanding of crowd behaviour thus requires a framework that captures these social, symbolic, and spatial entanglements; an integrative perspective that we have aimed to pursue in this study.

\subsection*{Relevance from today's perspective}
The Amritsar massacre stands as a cautionary tale not only of colonial brutality but also of the dangers inherent in dehumanising large groups that represent oppositional ideas and interests. The political opposition of the 19th century expressed its interests through its presence on the streets due to a lack of other possibilities. Thus, the perception of crowds as irrational, dangerous entities, a view rooted in 19th century crowd psychology, gradually expanded to political groups with unpreferred views \cite{garcia-nationalism_2015}. Although overt colonialism has receded, basic psychological mechanisms remain alarmingly relevant: fear of the ``mass,'' the temptation to view collective dissent as chaos, and the ease with which authority can escalate violence under the guise of preserving order. 
In an era of increasing social polarisation and radicalisation, recognising and resisting these mental patterns remains crucial. 

Modern political discourse continues to reflect these patterns. In 2020, during nationwide protests following the killing of George Floyd, President Donald Trump referred to Black Lives Matter protesters as ``thugs'' and warned, ``\textit{when the looting starts, the shooting starts}'' \cite{nyt_trump_looting_2020}.
Additionally, Trump urged governors to ``dominate'' protesters, suggesting that failure to do so would make them look like ``jerks'' \cite{vf_trump_bunker_2020}. During the same period, North Carolina Republican Representative Larry Pittman described the protesters as ``ignorant thugs,'' ``criminals,'' ``domestic terrorists,'' and ``vermin.'' He further stated that if protesters resisted and attacked police, they should be shot, asserting, ``This is war'' \cite{newsweek_senator_vermin_2020}.
In response to the 2019–2020 Hong Kong protests, Chinese state media labeled demonstrators as ``rioters'' and ``terrorists,'' framing the pro-democracy movement as a threat to national security \cite{bnn_china_hk_2019}. 
During the 2011–2013 Russian protests, state media and officials often portrayed demonstrators as foreign agents or traitors, suggesting that they were tools of external powers aimed at destabilising Russia \cite{nyt_putin_protests_2011}. 
In 2013, Turkish President Recep Tayyip Erdoğan labelled Gezi Park protesters as ``çapulcu'' (looters, marauders), a term that sought to delegitimise and dehumanise the demonstrators \cite{wiki_capulcu_2013, pomeps_erdo_gezi_2013}. These rhetorics framed the protestors as irrational and criminal, thus justifying harsh crackdowns and preemptive state violence.

These rhetorical strategies are not mere expressions of disdain: they serve functional roles in moral disengagement and the justification of state violence. Studies consistently show that dehumanisation reduces empathy, increases support for punitive action, and enables large-scale human rights violations \cite{haslam_dehumanization_2014, zlobina_back_2023}. When protestors are framed as ``mobs'', ``terrorists'' or ``vermin'', violence against them can be rebranded as a necessary security policy.

The memory of Amritsar, then, is not only historical, it is diagnostic. It reminds us that the crowd must not be treated as a faceless threat, but as a collection of individuals whose humanity must be affirmed, especially in moments of fear, unrest, and uncertainty.

\section*{Simulation results}
\label{sec:simulation}

We performed a series of simulations to assess the effects of spatial exposure, temporal progression, shielding, and population size on the collapse dynamics during the Jallianwala Bagh massacre scenario. All simulations were repeated ten times with different random seeds for statistical robustness.

\subsection*{Sensitivity to crowd size and crowding effect}
\begin{figure*}[htbp]
\centering

\subfloat[5\,000 pedestrians.\label{fig:time_series5k}]{
\includegraphics[width=0.32\textwidth]{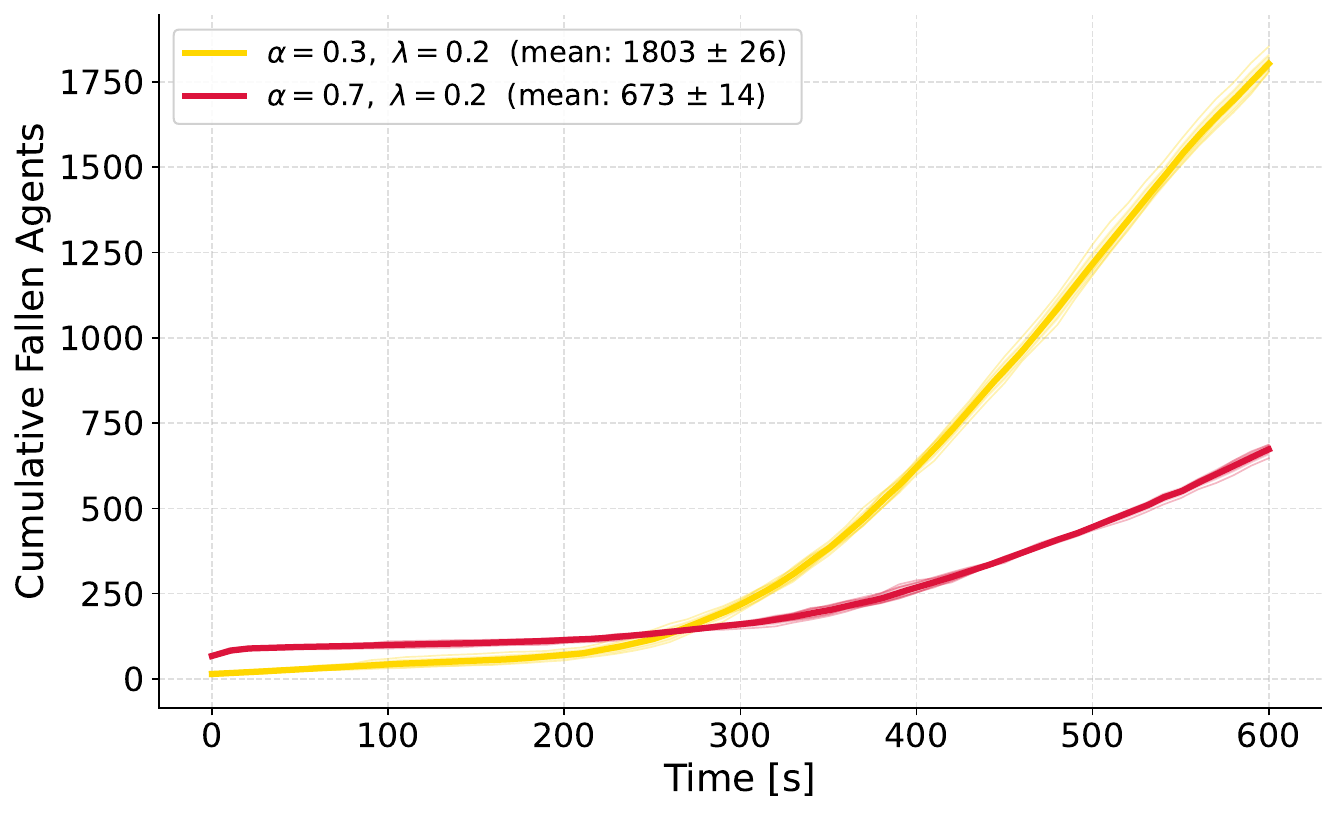}}
\hfill
\subfloat[10\,000 pedestrians.\label{fig:time_series10k}]{
\includegraphics[width=0.32\textwidth]{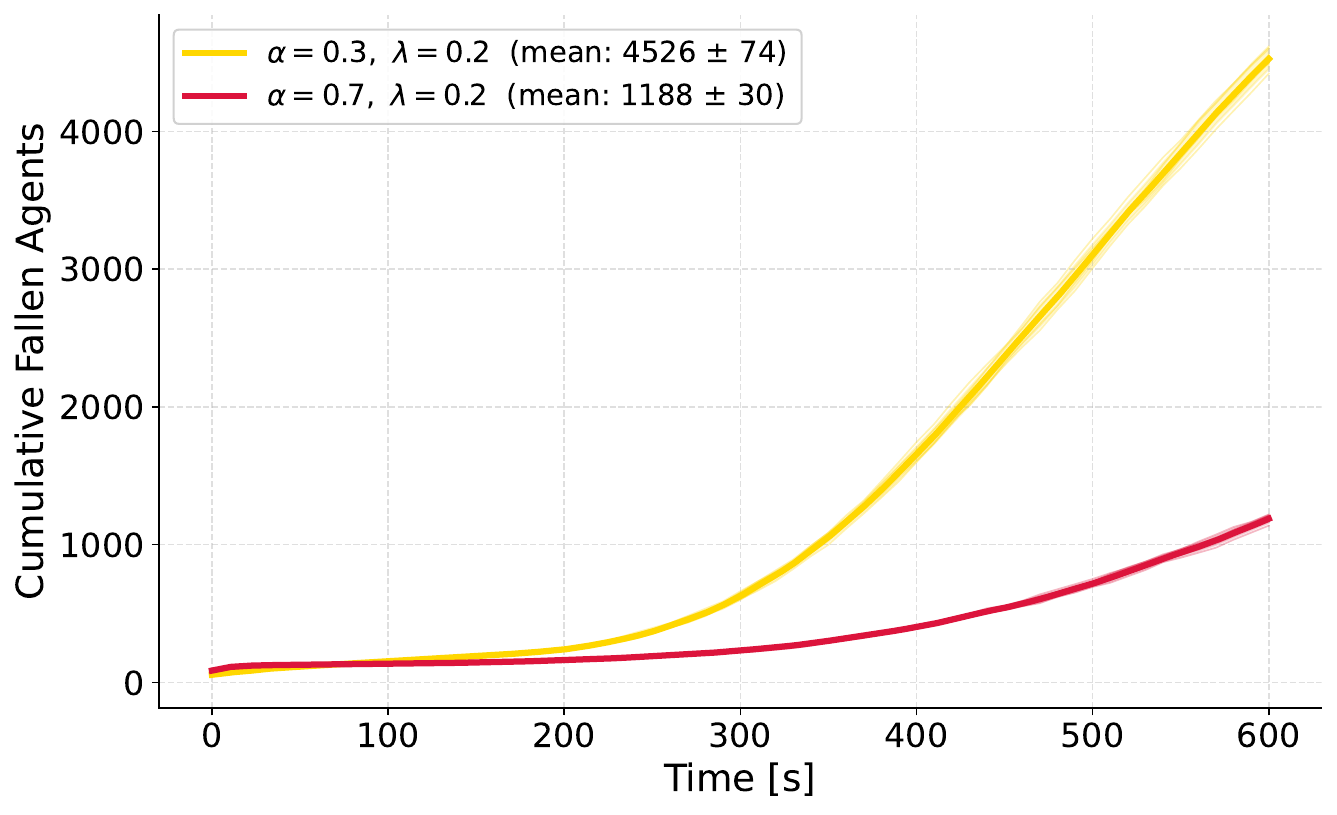}}
\hfill
\subfloat[15\,000 pedestrians.\label{fig:time_series15k}]{
\includegraphics[width=0.32\textwidth]{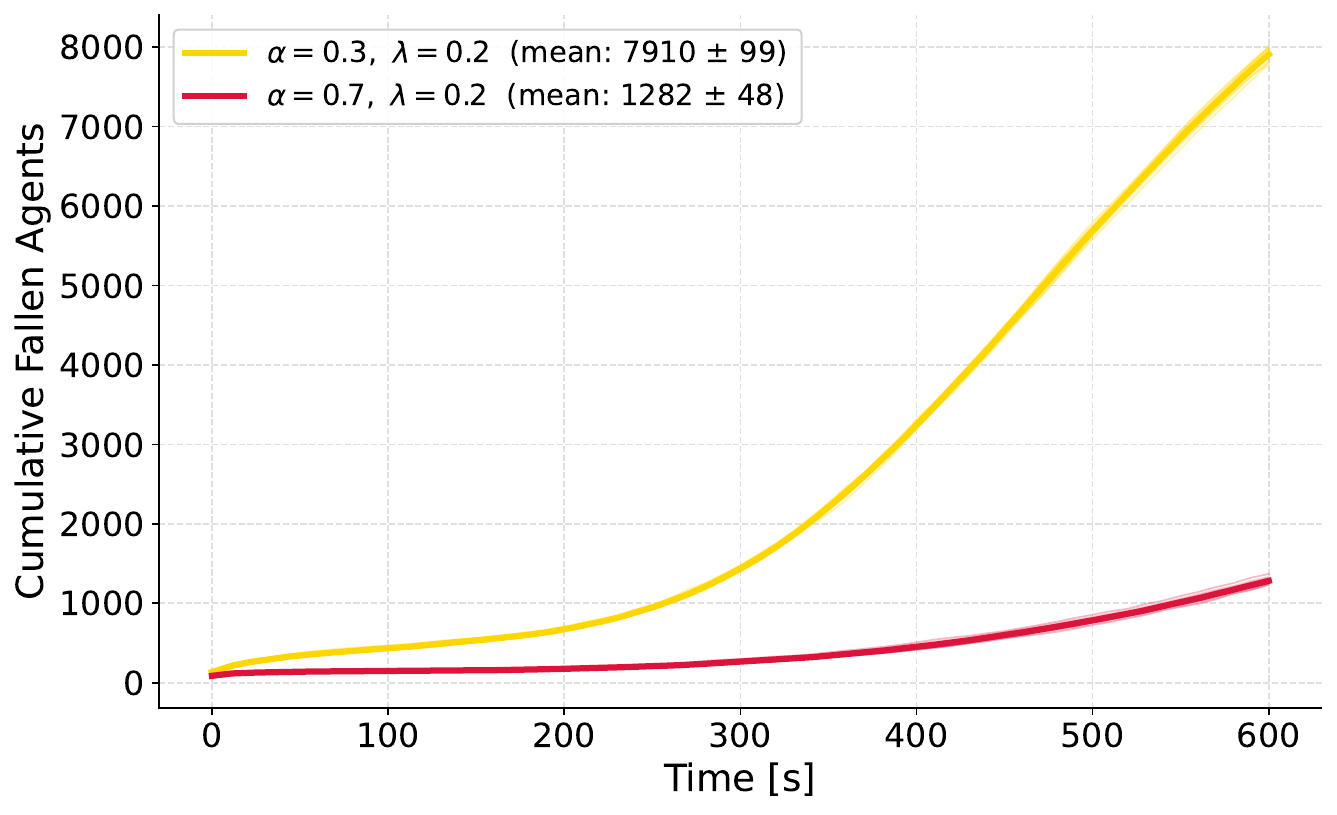}}

\caption{Cumulative time series of fallen people for different initial numbers of pedestrians for different $\lambda$ values.}
\label{fig:time_series}

\end{figure*}

We investigate how variations in crowd size affect the cumulative number of fatalities, given that historical estimates for the number of people present in the Bagh vary widely from around 5\,000 to more than 20\,000~\cite[p.~8]{Wagner2019}. 
Figure~\ref{fig:time_series} shows the evolution of cumulative fatalities over time for three different crowd sizes ($N = 5\,000$, $N = 10\,000$, and $N = 15\,000$) under two contrasting shielding scenarios: $\alpha = 0.3$ (representing clustered exposure) and $\alpha = 0.7$ (representing crowd shielding). 
For details on the dual role of the shielding effect, see Section \nameref{sec:crowd_shielding}.

For the smallest scenario (Figure~\ref{fig:time_series5k}), simulations with $\alpha = 0.7$ produce a mean of approximately 673 fallen pedestrians ($\pm$ 26), while for $\lambda = 0.3$, this number rises to about 1803 ($\pm$ 26). This corresponds to approximately 13\% and 36\% of the total population, respectively.
With $N = 10\,000$ (Figure~\ref{fig:time_series10k}), the relative difference becomes even more striking. Simulations with $\alpha = 0.7$ result in around 1188 ($\pm$ 30) fallen agents, or approximately 12\%, while $\alpha = 0.3$ yields about 4526 ($\pm$ 74), corresponding to roughly 45\% of the population.
In the densest scenario (Figure~\ref{fig:time_series15k}), the proportion of collapse again increases significantly. For $\alpha = 0.7$, the average number of fallen agents is around 1282 ($\pm$ 48), and for $\lambda = 0.7$, this climbs to about 7910 ($\pm$ 99), corresponding to approximately 9\% and 53\% of the crowd, respectively.


These results demonstrate that increasing the crowd size leads to a strongly non-linear rise in cumulative fatalities, highlighting how higher density amplifies local collapse propagation. 
The crowd shielding effect ($\alpha$) significantly modulates this outcome: when shielding is strong ($\alpha=0.7$), fewer agents are exposed to direct risk, resulting in markedly fewer casualties across all scenarios. 
However, as density increases, the protective effect of shielding diminishes in relative terms because the sheer number of people and restricted movement create conditions where collapse can cascade more easily. 
The shape of the time series suggests a critical threshold where local collapses trigger larger systemic failures, underlining how both crowd size and internal dynamics jointly govern the scale of the tragedy.

Overall, the consistently higher simulated fatalities, exceeding historical official records in all plausible scenarios, reinforce the view that the true scale of the event was likely underreported. 

\subsection*{Spatial distribution of fatalities}

\begin{figure*}[htbp]
\centering

\subfloat[5\,000 pedestrians.\label{fig:spatial_fatalities5k}]{
\includegraphics[width=0.32\textwidth]{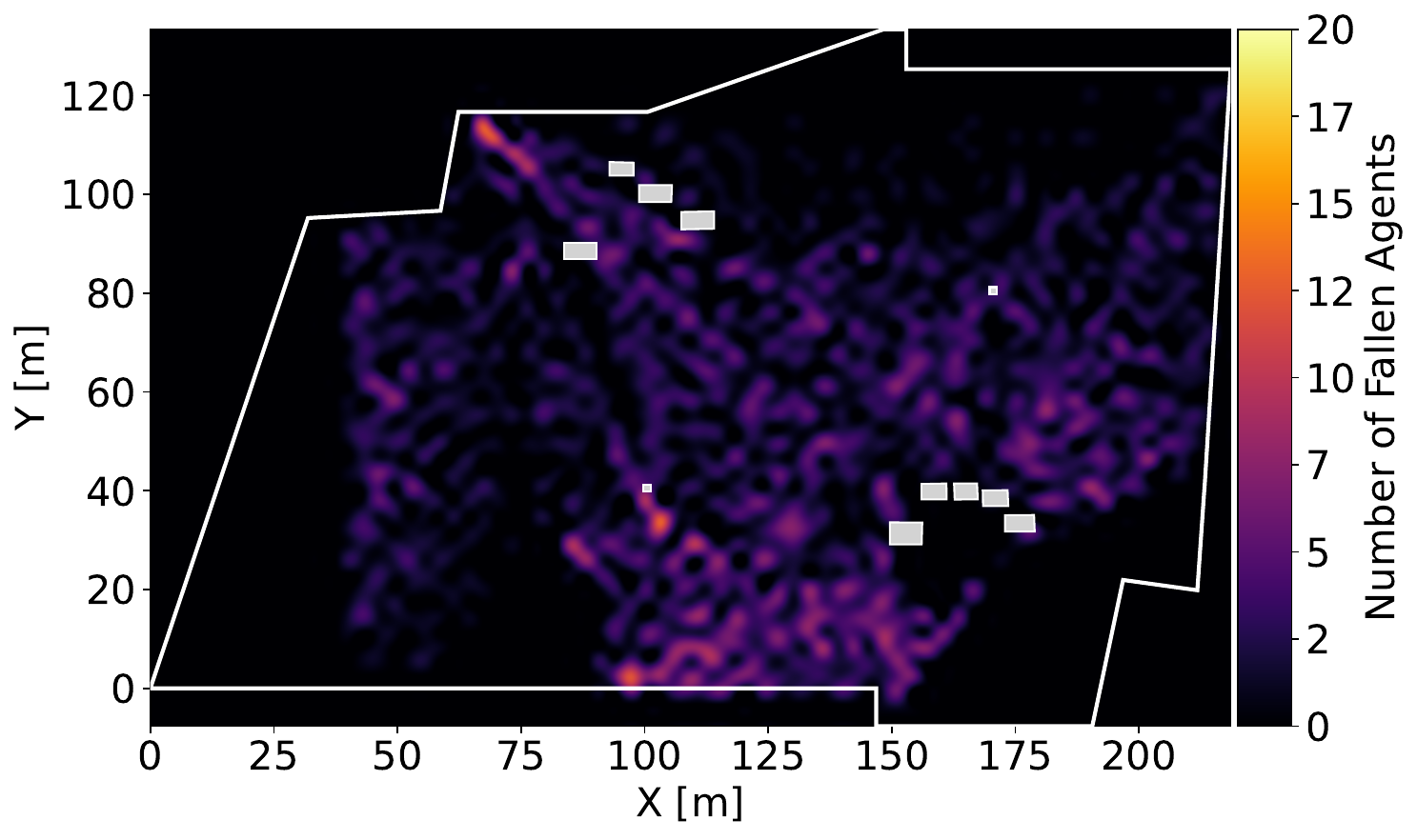}}
\hfill
\subfloat[10\,000 pedestrians.\label{fig:spatial_fatalities10k}]{
\includegraphics[width=0.32\textwidth]{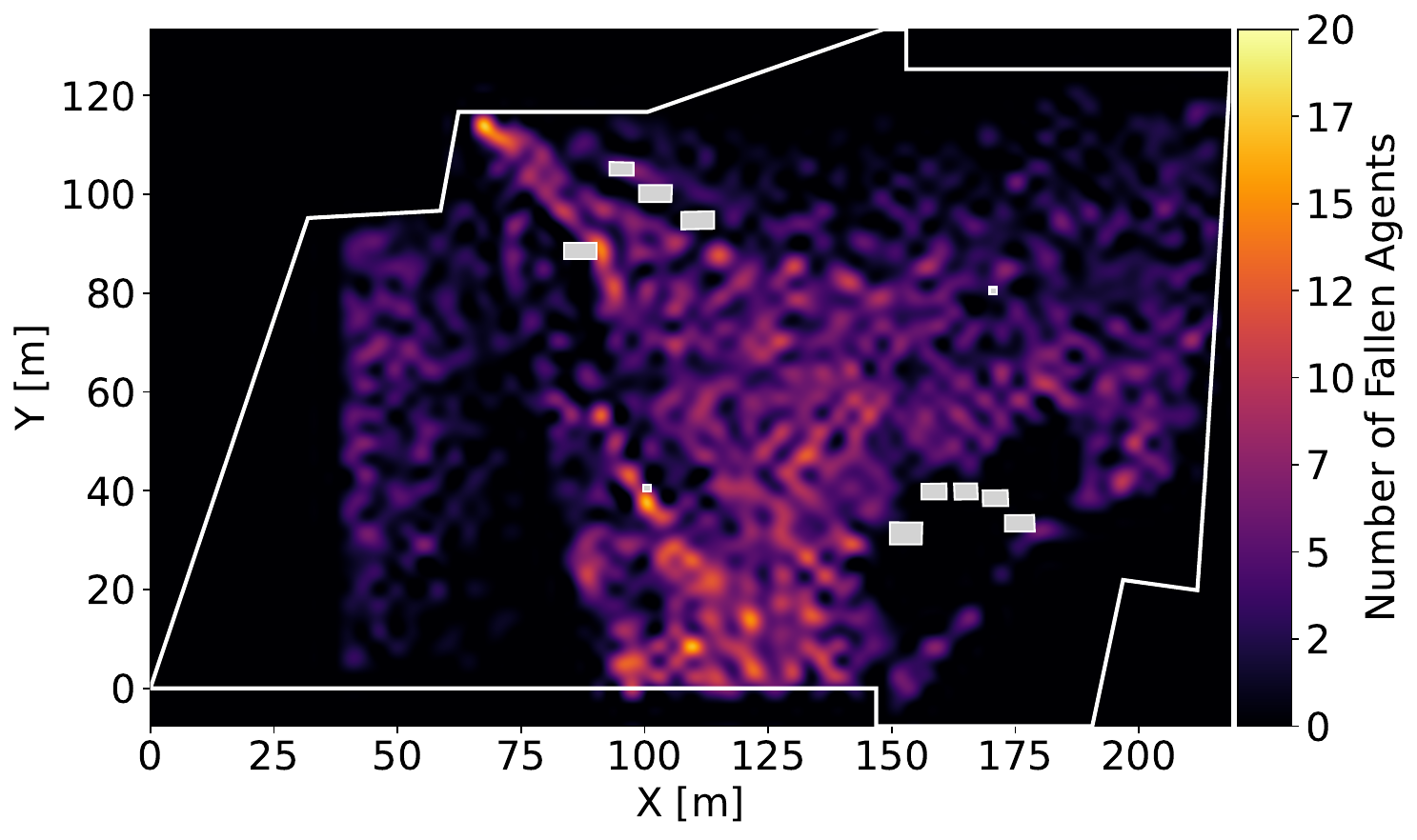}}
\hfill
\subfloat[15\,000 pedestrians.\label{fig:spatial_fatalities15k}]{
\includegraphics[width=0.32\textwidth]{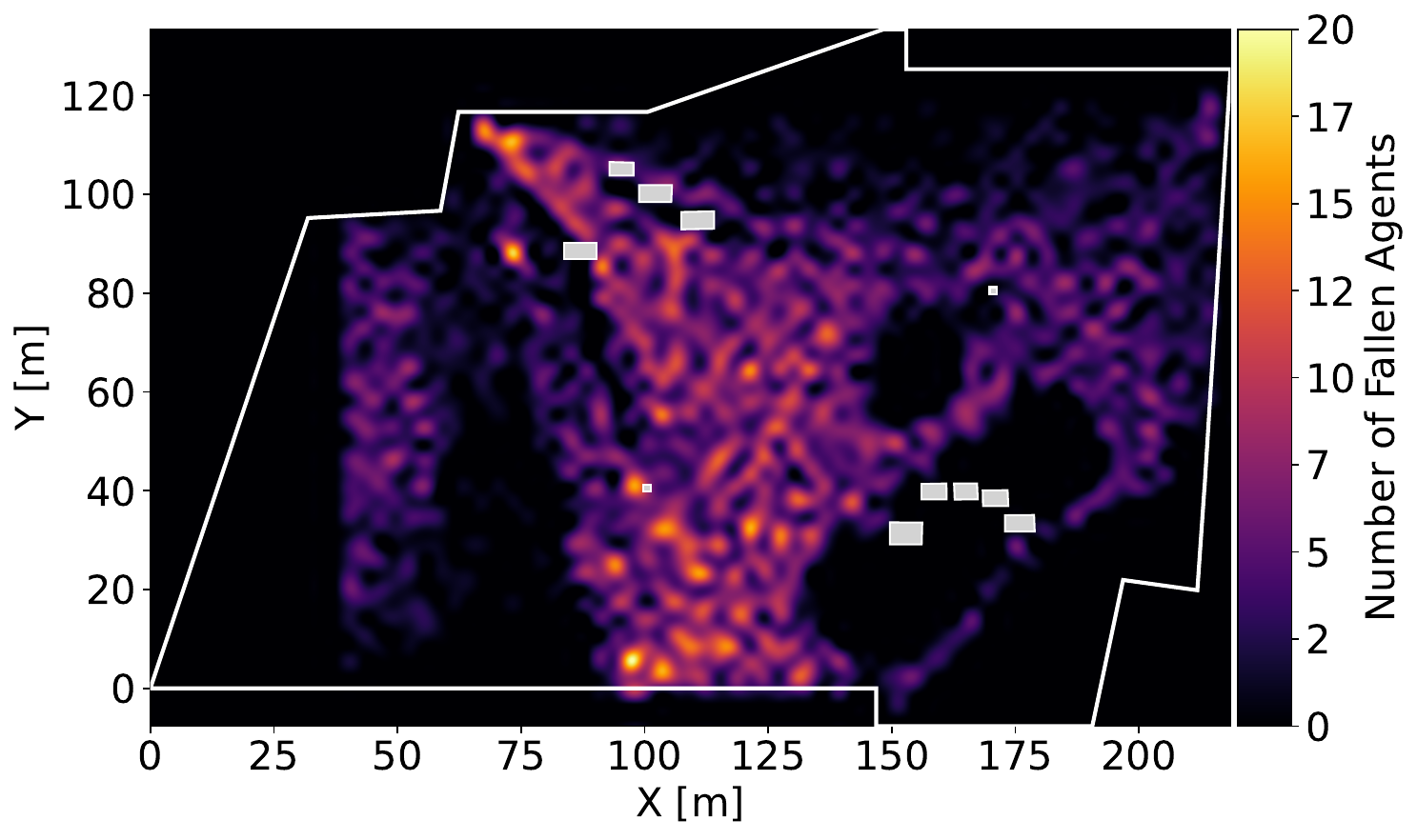}}

\caption{Summary of spatial distribution of fallen pedestrians for different numbers of pedestrians initially inside the Bagh ($\lambda = 0.5$).}
\label{fig:spatial_fatalities}

\end{figure*}

To understand where agents were most likely to collapse, we generated heat maps of the final positions of fallen agents, aggregated over ten simulation runs for each crowd size (Figure~\ref{fig:spatial_fatalities}). 
The spatial patterns reveal that the highest concentrations of fatalities do not occur directly at the shooting line (located around \( x = \SI{50}{\meter} \)), but rather deeper inside the Bagh, particularly in the central and lower right regions (\( x \approx \SIrange{100}{150}{\meter},\ y \approx \SIrange{0}{50}{\meter} \)).

This outcome reflects the emergent interplay of spatial exposure, local crowd density, and movement constraints. 
While the front line is exposed first to gunfire, it is also less congested, allowing some agents to escape or collapse without creating severe blockages. 
Deeper inside the garden, however, high densities and bottlenecks form as people attempt to flee, increasing exposure duration and leading to local collapse clusters. These clusters can block escape routes, resulting in further accumulation of fatalities in areas far from the initial threat.

As the total crowd size increases, these dynamics intensify. At \( N = 5\,000 \), fatalities remain relatively localized along a corridor extending inward from the entrance (Figure~\ref{fig:spatial_fatalities5k}).
For \( N = 10\,000 \), the affected region expands laterally, with new hotspots forming along the edges of the garden and around internal obstacles (Figure~\ref{fig:spatial_fatalities10k}), highlighting how structural features can amplify or redirect crowd flows. At \( N = 15\,000 \), nearly the entire lower half of the Bagh shows significant fatality densities, indicating widespread crowd compression and immobility. Notably, areas farthest from the shooting line still exhibit elevated fatalities due to secondary crowding effects (Figure~\ref{fig:spatial_fatalities15k}).

Obstacles play a dual role in this dynamic: in some areas, they create ``shadow zones'' where local shielding lowers fatalities, while in others, they become pinch points that exacerbate congestion and hinder movement, driving local collapse cascades. The visible arc-shaped and band-like clustering in the densest scenario suggests that immobility and chain reactions of falling agents can propagate far from the direct line of fire.

Overall, these findings demonstrate that spatial vulnerability during mass casualty events is shaped not just by direct line-of-sight exposure, but by the interaction of geometry, density, and collective crowd behaviour. Such dependencies are critical to capture, as they show how indirect effects like confinement and crowd compression can become equally, if not more, deadly than the direct threat itself.

\section*{Conclusion}

This work presents a social psychological interpretation and a novel simulation-based framework for reconstructing mass-casualty events in dense crowd scenarios, specifically applied to the case of the Jallianwala Bagh massacre. As Wani \cite{wani_hammer_2020} has noted, very few studies examined the Amritsar massacre within a broader societal context, and a comprehensive reappraisal of the events has only begun in recent years. In this regard, Wani situates the massacre within the framework of colonial history. Our approach, by contrast, seeks to contextualize the event through the lens of how crowds have been perceived as irrational entities. Emphasizing the historical framing of physical gatherings, we further discuss the contemporary implications for face-to-face assemblies and the ways in which state authorities respond to them. Regarding the reconstruction of the shooting in the Bagh and its potential outcomes, a complex model was introduced to account for factors at both the individual and collective levels. By integrating agent-based modelling with spatial-temporal exposure metrics, we have developed a tool capable of estimating collapse dynamics in the face of coordinated gunfire, under historical architectural constraints. Our findings suggest that with plausible crowd densities and shooting rates, casualty numbers may far exceed official historical accounts.

Beyond historical relevance, our results have broader implications for both forensic analysis and modern crowd safety. Typically, crowd simulations are designed to optimize event logistics such as ingress, egress, and evacuation in response to structural bottlenecks. However, our approach shifts the focus to scenarios where the crowd itself becomes the target, such as in mass shootings at public events, including recent tragedies at concerts, religious gatherings, and chrismas markets. 
These situations are qualitatively different from typical crowd disasters, as they involve intentional externally imposed threats rather than hazards that emerge organically from within the crowd.

Importantly, current pedestrian and evacuation models are not equipped to simulate these scenarios. They often do not account for dynamic exposure, local shielding, and irregular collapse patterns arising from targeted violence. Our study shows that modelling these dynamics can produce insights into not only how many people might be affected, but also where and when they are most vulnerable.

A major strength of this framework lies in its ability to serve as a forensic tool. In historical cases where records are incomplete, contested, or politically charged — as is the case with Amritsar — simulation can help challenge or corroborate reported outcomes. 
By grounding such simulations in physically and behaviorally plausible assumptions, while incorporating established historical facts, it becomes possible to evaluate whether official narratives align with the spatial constraints and fundamental mechanics of crowd movement.

In this sense, the model acts as a form of retrospective validation, capable of providing a ``scientific estimate'' when eyewitnesses are gone and photos are rare.

Several promising avenues remain for future work. 
First, incorporating additional behavioural responses, group cohesion, or protective actions would enrich the realism of the simulation. 
Second, we plan to integrate confined escape zones (such as the well in Jallianwala Bagh) to better capture structural vulnerabilities. 
Third, our approach could be extended to simulate contemporary high-risk events, aiding in both prevention planning and post-incident analysis. 
Finally, developing public-facing tools from such models can empower communities to critically engage with contested historical accounts, especially in contexts where state violence or crowd mismanagement are underreported.

\section*{Acknowledgements}
The authors would like to thank Kim A. Wagner for discussions and introduction to this historical event.
\section*{Funding Declaration}
M.C. gratefully acknowledges financial support from the BMBF project Croma-Pro, grant number DB002398.

\section*{Author contributions} 
K.K. and E.Ü. conceptualized and developed the historical and social psychological framework of the study.
M.C. designed and implemented the model.

All authors contributed to the interpretation of the results, engaged in critical discussion, and collaboratively wrote the final manuscript.

\section*{Code availability}
The model implementation is publicly available on Zenodo \cite{mohcine_chraibi_2025_16264825}. It was developed using \textsc{JuPedSim} \cite{jupedsim}, an open-source framework tailored for simulating pedestrian dynamics in complex environments.


\clearpage 

%
\bibliography{reference} 
\bibliographystyle{sciencemag}

%
%
%
%
%
\end{document}